\documentclass{article}

\usepackage[english]{babel}

\usepackage[letterpaper,top=2cm,bottom=2cm,left=3cm,right=3cm,marginparwidth=1.75cm]{geometry}

\usepackage{amsmath}
\usepackage{graphicx}
\usepackage[colorlinks=true, allcolors=blue]{hyperref}
\usepackage{amsmath,amssymb,amsfonts}%
\usepackage{amsthm}%
\usepackage{mathrsfs}%
\usepackage{xcolor}%
\usepackage{textcomp}%
\usepackage{manyfoot}%
\usepackage{booktabs}%
\usepackage{algorithm}%
\usepackage{algorithmicx}%
\usepackage{listings}%
\usepackage{hyperref}
\usepackage{graphicx}
\usepackage{subcaption}
\usepackage{multirow}
\usepackage{soul}
\usepackage{lineno}
\usepackage{authblk}

\title{Optimising Underwater Neutrino Telescopes for All-Flavour Point Source Sensitivity}

\author[1]{Iwan~Morton-Blake}
\author[1]{Fuyudi~Zhang}
\author[1]{Qichao~Chang}
\author[1]{Shuhua~Hao,}
\author[1]{Weilun~Huang}
\author[1,2,3]{Hualin~Mei}
\author[1]{Wei~Tian}
\author[1]{Yingwei~Wang,}
\author[1,2,3]{Xin~Xiang}
\author[1,2,3]{Donglian~Xu\thanks{Corresponding author: \texttt{donglianxu@sjtu.edu.cn}}}

\affil[1]{State Key Laboratory of Dark Matter Physics, Tsung-Dao Lee Institute, Shanghai Jiao Tong University, Shanghai 201210, China}
\affil[2]{School of Physics and Astronomy, Shanghai Jiao Tong University, Key Laboratory for Particle Astrophysics and Cosmology (MoE), Shanghai Key Laboratory for Particle Physics and Cosmology, Shanghai 200240, China}
\affil[3]{Hainan Research Institute, Shanghai Jiao Tong University, Hainan 572024, China}

\begin{document}
\maketitle

\begin{abstract}

High-energy neutrino astronomy has advanced rapidly in recent years, with IceCube, KM3NeT, and Baikal-GVD establishing a diffuse astrophysical flux and pointing to promising source candidates. These achievements mark the transition from first detections to detailed source studies, motivating next-generation detectors with larger volumes, improved angular resolution, and full neutrino-flavour sensitivity. 
We present a performance study of large underwater neutrino telescopes, taking the proposed TRIDENT array in the South China Sea as a case study, with a focus on comparing the performance of various detector configurations against the TRIDENT baseline design. Both track-like events primarily from muon neutrinos, which provide precise directional information, and cascade events from all flavours, which offer superior energy resolution, diffuse-source sensitivity, and all-sky flavour coverage, are included to 
achieve a balanced performance across source types. The time to discover potential astrophysical sources with both track- and cascade-like events is used as the figure of merit to compare a variety of detector design choices. Our results show that, for a fixed number of optical modules, simply enlarging the instrumented volume
does not inherently lead to improved performance, while taller strings can provide modest gains across all detector channels, within engineering constraints. Distributing dense clusters of strings over a large volume is found to generally worsen discovery potential compared to the baseline layout. Finally, the optical properties of the sea-water emerge as the key factor dictating the optimisation of detector layout, highlighting the need for in-situ measurements and early deployment of optical modules to guide the final array configuration.

\end{abstract}

\section{Introduction}

High-energy neutrinos provide a unique probe of the Universe’s most extreme environments. As neutral, weakly interacting particles, 
\textcolor{black}{they propagate almost undeflected and with negligible attenuation over cosmological distances, offering direct insight into hadronic acceleration mechanisms and serving as messengers from dense astrophysical environments that are opaque to electromagnetic radiation\cite{Murase:2013rfa}.} Since Markov’s 1960s proposal for large-volume Cherenkov detectors~\cite{markov}, neutrino astronomy has advanced from concept to an experimental reality.

The IceCube Neutrino Observatory has established the presence of a diffuse astrophysical neutrino flux~\cite{IceCube:2013low}, and provided mounting
evidence for specific sources, including a potential association with the blazar TXS~0506+056~\cite{IceCube_TXS_flares:2018}, the identification of active galaxy NGC~1068 as a compelling point source~\cite{IceCube:2022der}, 
\textcolor{black}{and the detection of high-energy neutrinos from the Galactic plane}~\cite{IceCube:2023ame}. After more than a decade of operation, IceCube has firmly established the existence of an astrophysical neutrino flux and begun to probe its spectral shape \cite{IceCube:2024fxo, IceCube:2025dlr}, sky distribution \cite{IceCube:2025ary}, and potential source classes \cite{McDonough:2023ngk}. However, many results remain limited by statistical precision and by the ability to localise sources. This has motivated the development of complementary instruments in the Northern Hemisphere, with improved angular resolution, flavour sensitivity, and sky coverage.

Water-based neutrino telescopes have begun to extend this reach. KM3NeT has announced the observation of an ultra-high-energy neutrino with an estimated energy of $\sim$220PeV~\cite{KM3NeT:2025npi}, the highest-energy neutrino ever detected, while the Baikal-GVD array recently reported a $5\sigma$ detection of the diffuse flux using cascade events~\cite{Baikal-GVD:2022fis,Baikal-GVD:2025update}. Together, these facilities mark the transition from first detections to wider coverage and higher energies. Yet, significant challenges remain: the sample of identified astrophysical neutrinos is still small, the number of confirmed sources is limited, and sensitivity to the neutrino flavour composition, especially for tau neutrinos, offers scope for enhancement.

\textcolor{black}{TRIDENT is designed to address these challenges. The experiment is a proposed deep-sea neutrino telescope to be deployed in the South China Sea~\cite{TRIDENT:2022hql}, with primary physics goals that include: (1) the rapid discovery and identification of astrophysical neutrino sources, and (2) improved all-flavour sensitivity, enabling precise measurements of the neutrino flavour ratio and searches for new physics~\cite{IceCube:2021tdn, Arguelles_snowmass:2022}.} These objectives directly support the broader aim of probing the acceleration mechanisms of high-energy cosmic rays.

To achieve TRIDENT’s ambitious goals and fully exploit its potential, it is essential to systematically evaluate the telescope’s design and performance across different event types and source scenarios. In this work, we use the dedicated TRIDENT simulation framework, hereafter referred to as \textit{TRIDENTSim}, as a case study to conduct a comprehensive performance study aimed at informing the optimal geometry of large-scale deep-sea neutrino telescopes.

Rather than focusing solely on neutrino event rates or track angular resolution, we adopt a joint approach that considers both track-like and cascade-like event channels (described further in section~\ref{sec:events_in_trident}), ensuring sensitivity across the full sky and to all neutrino flavours. Track events, primarily from muon neutrinos, provide the highest directional precision, while cascades, produced by electron neutrinos, tau neutrinos, and neutral-current interactions of all flavours, contribute essential coverage in regions and energy ranges where track statistics are limited~\cite{IceCube_extended}. The analysis spans a representative range of astrophysical neutrino source scenarios, from hard ($E^{-2}$) to soft ($E^{-3}$) energy spectra, ensuring robustness to source model uncertainties. 
While TRIDENTSim serves as a case study, the methodology and conclusions are broadly applicable to the design of future high-energy neutrino observatories.

We explore variations in string spacing, string height, and clusters of strings, as well as the influence of deep-sea optical properties, particularly the attenuation length. The results show that optimal performance does not arise from maximising detector volume alone 
when the number of optical sensor is fixed; instead, it requires a balance between geometry and the optical environment. Indeed, moderate changes in attenuation length can have impacts comparable to—or larger than—geometry changes, making site characterisation a critical design driver. This underscores the importance of TRIDENT’s first ten string deployments, named Phase-I, as both a site survey, technology and physics demonstrator, guiding the final layout for the full array and informing future large-scale neutrino telescope designs.

\subsection{The TRIDENT Detector} 
As a case study for undersea neutrino telescope optimisation, we consider the TRIDENT detector concept in detail. The planned full array is expected to be deployed at a depth of 3.5~km in the South China Sea, made up of $\sim$1000 vertical strings anchored to the seabed, each instrumented with dozens of hybrid Digital Optical Modules (hDOMs) \cite{Hu:2021jjt, hDOM_34inch}. Each hDOM integrates multiple high–quantum efficiency photomultiplier tubes (PMTs) and silicon photomultiplier (SiPM) arrays into a pressure-resistant glass sphere, providing wide angular coverage, nanosecond timing, and local coincidence triggering~\cite{Wang:2025wqv, Zhi:2024dmr}. Strings will be distributed over several cubic kilometres, with geometry tuned for balanced performance in both track-like and cascade-like channels. The Phase-I detector instrumentation, comprising of approximately ten strings, will validate the full-chain of mechanical, electronic, data acquisition and calibration systems under deep-sea conditions ~\cite{T-REX:2024qfj, junhong_icrc2025, yangyong_icrc2025}, demonstrate neutrino detection and reconstruction and their separation from cosmogenic backgrounds, along with in-situ measurements of water optical properties and ambient backgrounds from radioactivities and bioluminescences. These results will feed directly into the geometry optimisation presented in this work.

\section{Event Topologies in TRIDENT}
\label{sec:events_in_trident}

Neutrino interactions in large-scale Cherenkov detectors such as TRIDENT produce two primary classes of observable event topologies: \textit{tracks} and \textit{cascades}, corresponding to different neutrino flavors and interaction types. Tau-neutrino charged-current (CC) interactions can produce characteristic double-cascade events; however, given their limited statistics, they are treated as cascades in this work.

\paragraph{Track-like events} are primarily produced in $\nu_{\mu}$ CC interactions, generating energetic muons that can travel kilometres through water. These long tracks provide the best handle on the incoming neutrino direction and form the basis of pointing sensitivity in neutrino telescopes. A major background for these events is atmospheric muons produced in cosmic ray interactions in the atmosphere~\cite{IceCube:2020wum}, which dominate the down-going event rate. As a result, astrophysical $\nu_{\mu}$ searches primarily rely on up-going tracks to suppress this background. Although veto strategies for down-going muons have been developed~\cite{Gaisser:2014bja}, they are not considered in this work.

Long muon tracks may also exit the detector volume, depositing only a portion of their energy within the instrumented region. This limits the precision of energy reconstruction for track events. In addition, atmospheric muon neutrinos produce an irreducible background that overlaps in topology with astrophysical tracks. In astrophysical source searches, improved angular resolution is the primary tool to mitigate contamination and enhance discovery potential.

\paragraph{Cascade-like events} arise primarily from $\nu_e$ and low-energy $\nu_\tau$ CC interactions, as well as neutral-current (NC) interactions of all flavors. These events produce \textcolor{black}{localized} energy depositions, typically {on the scale of meters, resulting in a more point-like Cherenkov light distribution compared to elongated tracks. This limits their angular resolution. However, the long scattering lengths of light in water still allow for moderate directional reconstruction. For extended sources that are intrinsically diffuse, such as sources within the Galaxy, extremely precise angular resolution is less critical. In these cases, cascades can provide a significant boost in sensitivity \cite{IceCube_extended}.

Cascades also provide advantages in source searches, as their energy is typically deposited entirely within the detector, leading to better energy resolution than for tracks. They also suffer reduced contamination from atmospheric muons and neutrinos. These features, along with full-sky coverage and sensitivity to all flavors, make cascades a critical component of modern neutrino telescope designs. Their inclusion enables balanced sensitivity across flavors and sky regions, and motivates optimisation strategies that consider both event topologies.

\vspace{0.5cm}

\textcolor{black}{TRIDENT is designed to efficiently and accurately reconstruct both track-like and cascade-like events. Dense arrangements of hDOMs vertically on each string allows for high photon detection efficiency and} precise timing-based reconstruction of cascade vertices and deposited energy, while extended lengths between each string yields a large instrumented volume, aiding high-energy muon detection and reconstruction.
The ability to reconstruct both topologies is crucial for maximising discovery potential across a wide range of potential source energy spectra, declinations, and flavor compositions. In this study, we evaluate the detector's discovery potential separately for track and cascade channels, under varying geometric configurations and optical properties. This dual-channel approach enables a more complete assessment of the detector’s capability to rapidly discover sources and maintain sensitivity to all neutrino flavors.

\begin{figure}[!ht]
    \centering
    \begin{subfigure}{0.48\textwidth}
        \includegraphics[height=2.8cm,width=\textwidth]{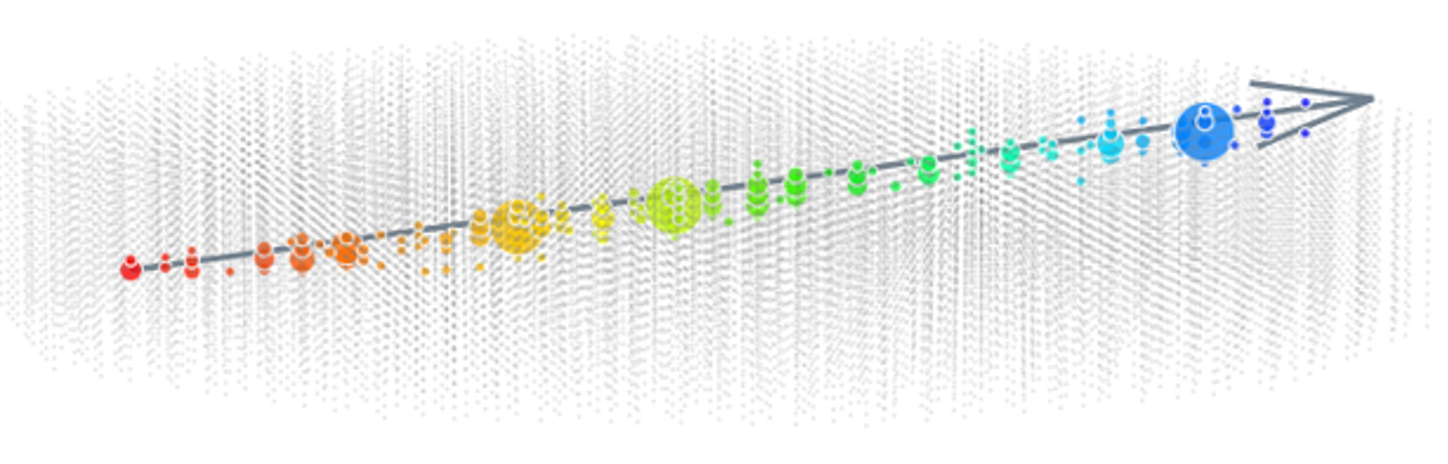}
        \subcaption{}
        \label{subfig:sim_track}
    \end{subfigure}
    \centering
    \begin{subfigure}{0.48\textwidth}
        \includegraphics[height=2.8cm, width=\textwidth]{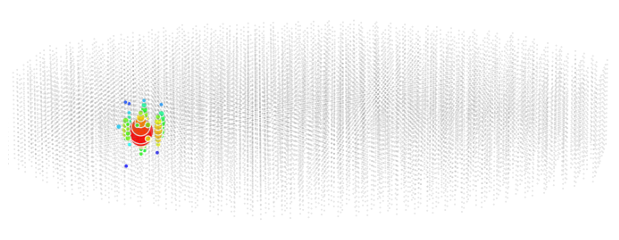}
        \subcaption{}
        \label{subfig:sim_cas}
    \end{subfigure}
    \caption{Visualization of simulated neutrino interactions using the TRIDENTSim framework, showing an example $\nu_{\mu}$-CC track event (a) and $\nu_{e}$-CC cascade event (b). Coloured spheres and their radii measure the number of hits detected on a given hDOM, where the colour scale indicates the time of detected photons, where red corresponds to early times and blue to late.}
    \label{fig:Sim_trackcascade}
\end{figure}

\section{TRIDENT Simulation Framework and Detector Design}
\label{sec:detector_simulation}

The sensitivity of a high-energy neutrino telescope to astrophysical sources depends on its geometrical configuration and the resulting optical photon detection efficiency. This section outlines the various detector geometry choices considered in this study, where the TRIDENTSim analysis chain \textcolor{orange}{is} used to simulate neutrino interactions and account for photon detection and detector response, with event reconstruction methods applied for both track- and cascade-like topologies.

\subsection{Detector Simulation Framework: TRIDENTSim}

To evaluate the physics sensitivity of each detector configuration, we use TRIDENTSim, a dedicated full-chain simulation framework based on \textsc{CORSIKA8}~\cite{CORSIKA8:2018}, \textsc{Pythia8}~\cite{Bierlich:2022pfr}, \textsc{CRMC}~\cite{crmc} and \textsc{Geant4}~\cite{GEANT4:2002zbu}. 

\paragraph{Neutrino generation and interaction:} Neutrino fluxes for all flavours (\( \nu_e, \nu_\mu, \nu_\tau \)) are generated and propagated through the Earth. Deep inelastic scattering interactions are forced in the vicinity of the detector. Final-state particles and showers are simulated with \textsc{Pythia8}, while secondary muon tracks and electromagnetic/hadronic cascade showers are propagated with \textsc{CRMC}/\textsc{Geant4}.

\paragraph{Cherenkov light propagation:} Charged particles produce Cherenkov photons, which are produced and tracked using \textsc{Geant4} and the \textsc{OptiX} ray-tracing engine~\cite{Opticks:2019}, accounting for absorption and scattering effects based on optical parameters measured at the TRIDENT site~\cite{TRIDENT:2022hql, camera_icrc2025}.


\paragraph{Photon detection:} Photon propagation within the hDOMs is simulated with detailed \textsc{Geant4} models, including PMT quantum and collection efficiencies, transit time spread, along with glass/gel interfaces and the PMT/SiPM support structure~\cite{hDOM_34inch}. 
\textcolor{black}{Although SiPMs offer measurable benefits, as discussed in~\cite{hDOM_34inch}, their design and implementation in reconstruction methods are an ongoing area of study, and are therefore not included in this work.}

Example track and cascade events simulated in the TRIDENTSim framework can be seen in Fig.~\ref{fig:Sim_trackcascade}, highlighting the significant differences in energy deposition possible for each interaction type within the detector.

\subsection{Detector Geometry Variation}

\begin{figure}[!ht]
    \centering
    \begin{subfigure}{0.3\textwidth}
        \centering
        \includegraphics[width=\linewidth]{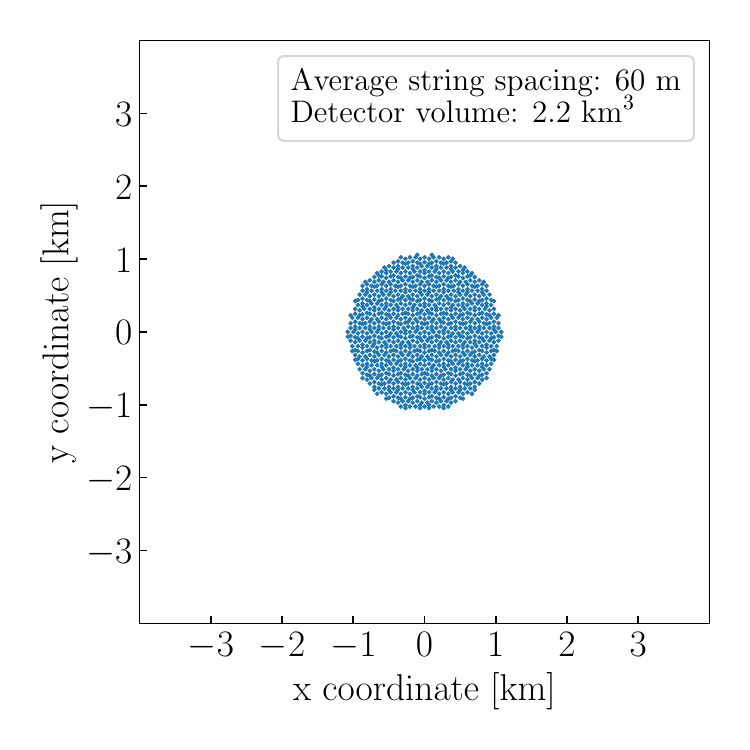}
        \subcaption{}
    \end{subfigure}
    \begin{subfigure}{0.3\textwidth}
        \centering
        \includegraphics[width=\linewidth]{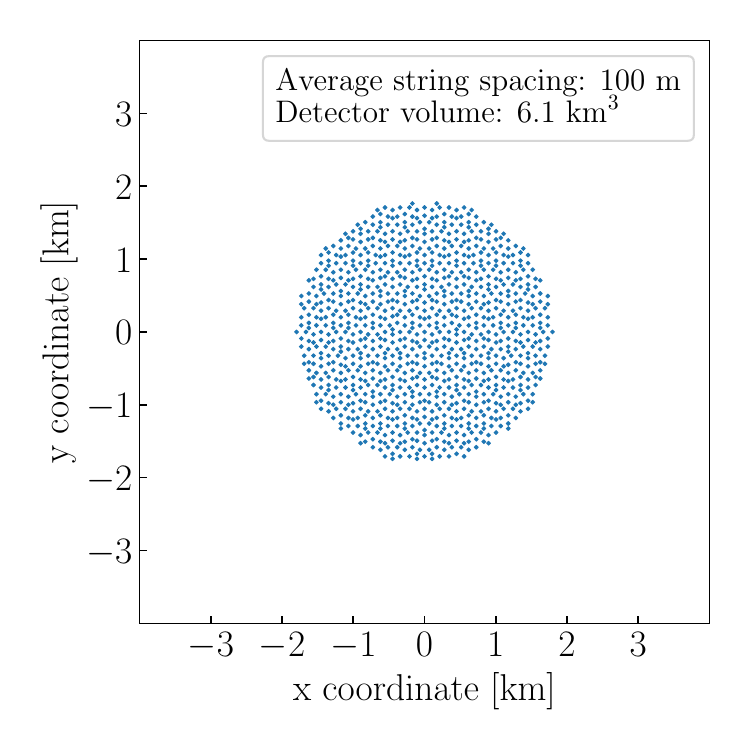}
        \subcaption{}
    \end{subfigure}
    \begin{subfigure}{0.3\textwidth}
        \centering
        \includegraphics[width=\linewidth]{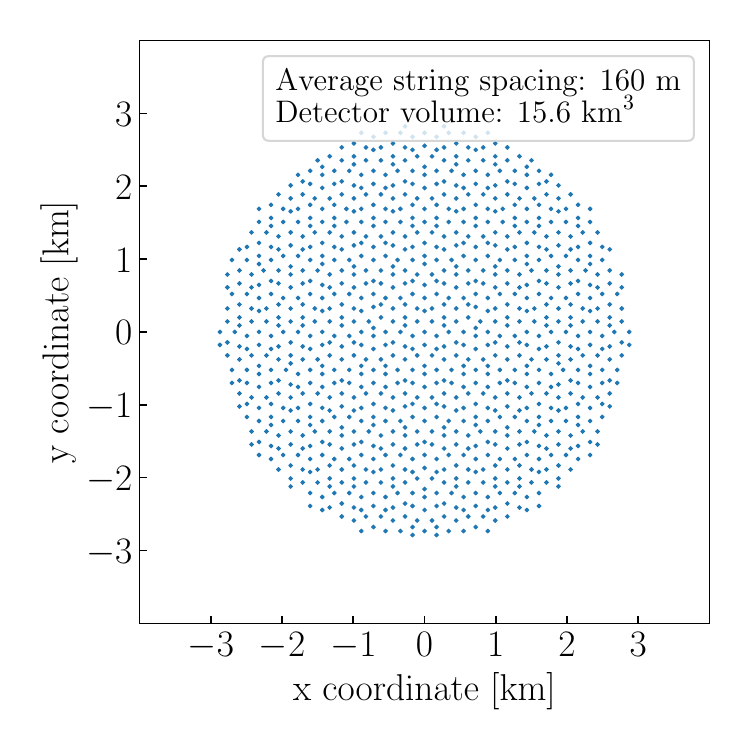}
        \subcaption{}
    \end{subfigure}
        \begin{subfigure}{0.3\textwidth}
        \centering
        \includegraphics[width=\linewidth]{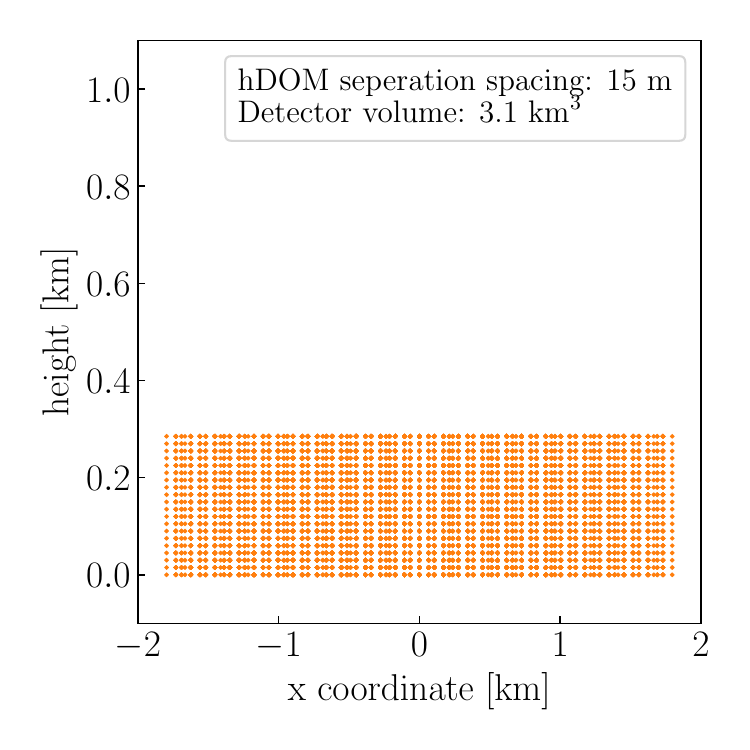}
        \subcaption{}
    \end{subfigure}
    \begin{subfigure}{0.3\textwidth}
        \centering
        \includegraphics[width=\linewidth]{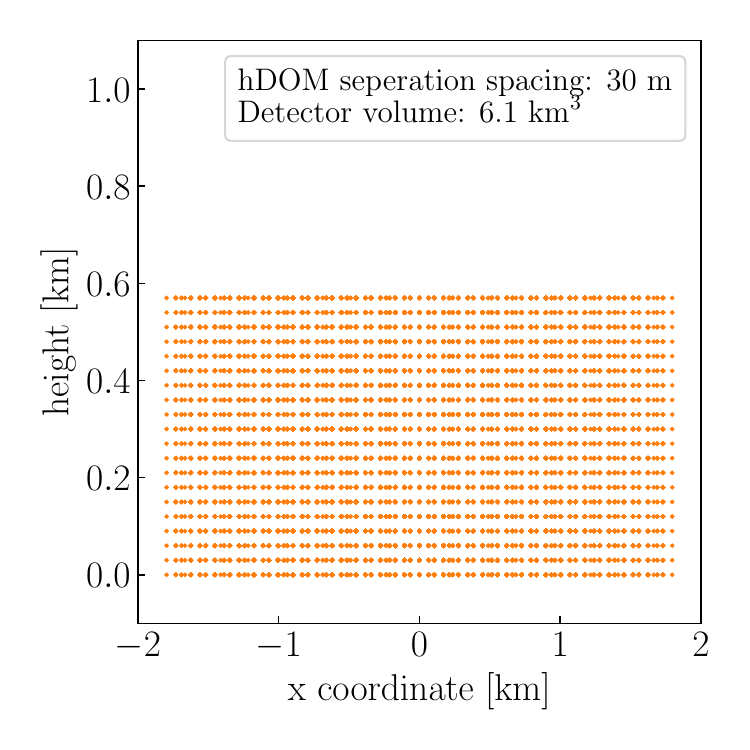}
        \subcaption{}
    \end{subfigure}
    \begin{subfigure}{0.3\textwidth}
        \centering
        \includegraphics[width=\linewidth]{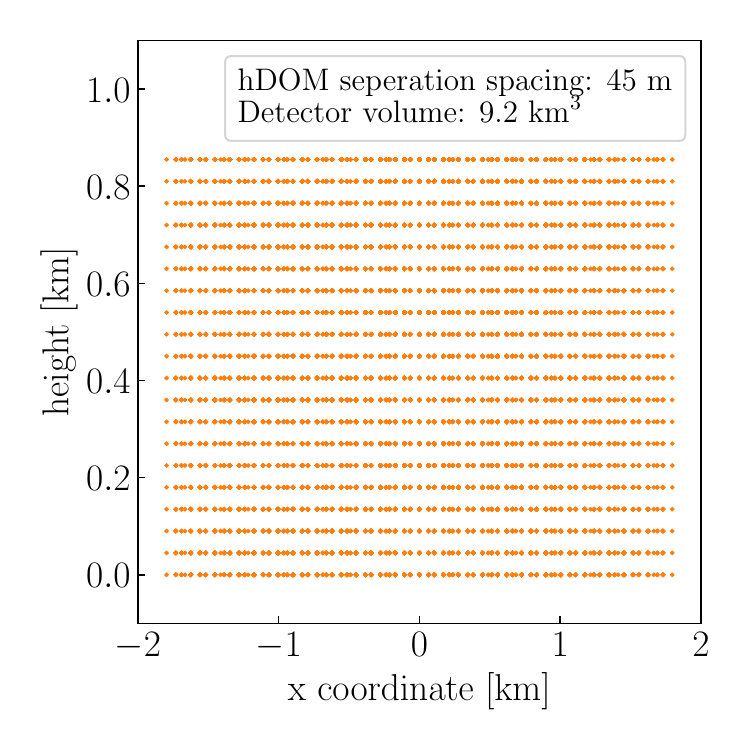}
        \subcaption{}
    \end{subfigure}
    \begin{subfigure}{0.3\textwidth}
        \centering
        \includegraphics[width=\linewidth]{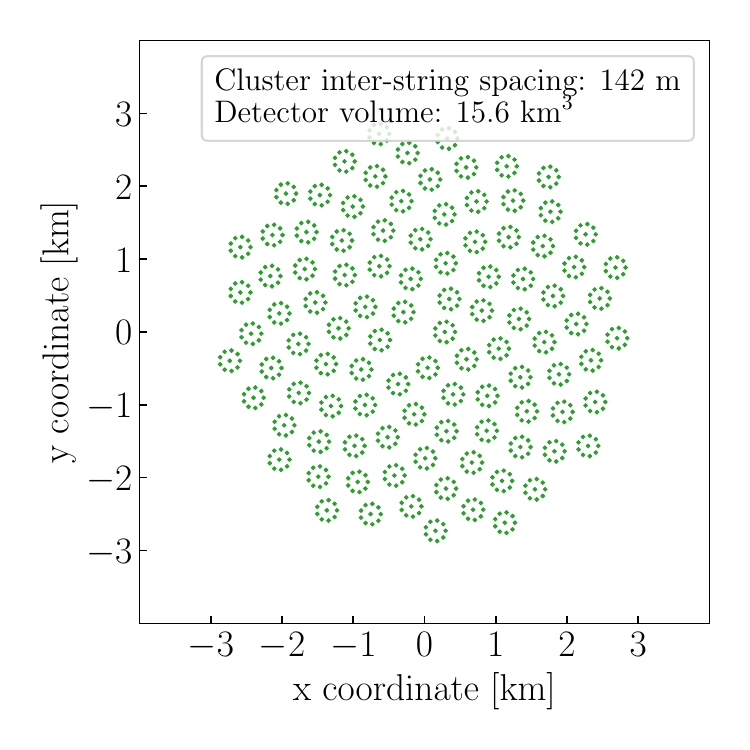}
        \subcaption{}
    \end{subfigure}
    \begin{subfigure}{0.3\textwidth}
        \centering
        \includegraphics[width=\linewidth]{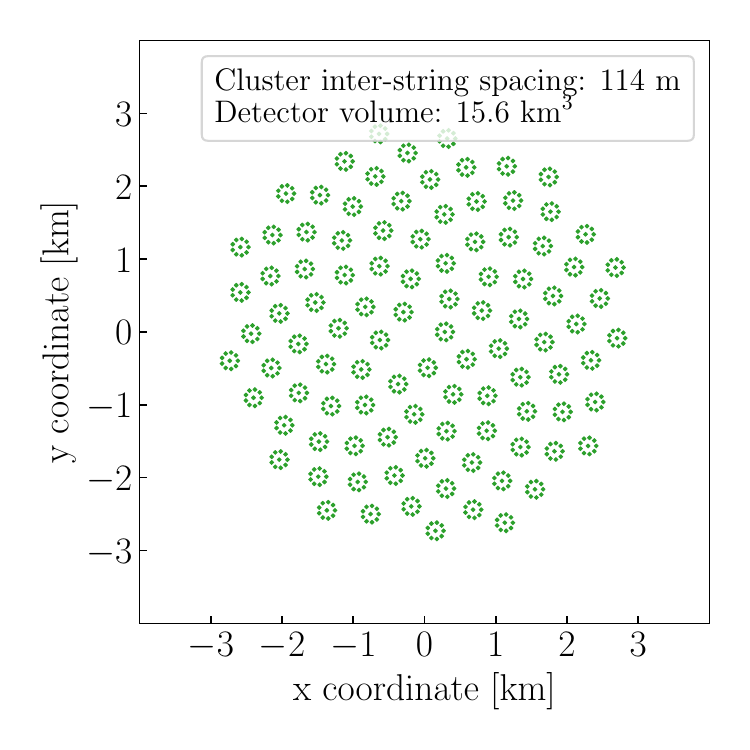}
        \subcaption{}
    \end{subfigure}
    \begin{subfigure}{0.3\textwidth}
        \centering
        \includegraphics[width=\linewidth]{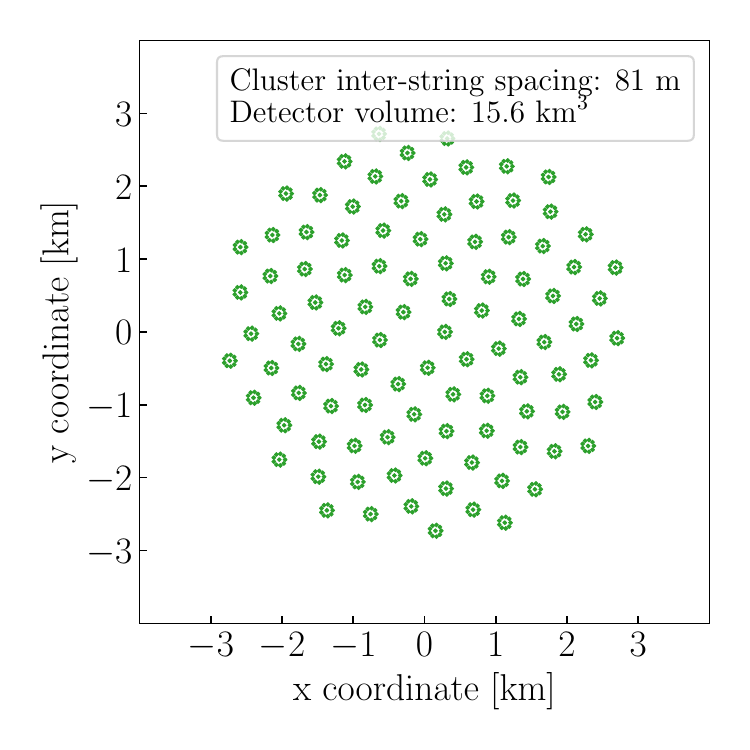}
        \subcaption{}
    \end{subfigure}
    \caption{ Example detector geometries tested in this study, where each point represents an hDOM. (a - c) \textit{String spacing:} top-down views of TRIDENT detector geometries, for string-to-string distances from 60 to 160 m, adhering to a Penrose tiling distribution of strings. (d - f) \textit{String spacing:} side-on views of detector geometries varying hDOM vertical spacing, where each string contains 20 hDOMs, separated by 15, 30 and 45 m. (g - i) \textit{String clustering:} Varied levels of string clustering, where the radius of the entire instrumented volume coincides with the largest string spacing detector is seen in (c). Figures (b) and (e) correspond to the reference TRIDENT geometry used as a comparison point reference in sections \ref{sec:reconstruction} and \ref{sec:point_source_discovery potential}.}
    \label{fig:DetGeoStringSpacing}
\end{figure}

The TRIDENT detector consists of vertical strings deployed in the deep ocean. For a fixed number of 1000 strings and 20 hDOMs per string, we study the impact of geometry on physics performance across three primary categories:

\paragraph{String spacing:} The horizontal spacing between strings determines the total instrumented volume and influences photon collection efficiency for events between widely spaced strings. Six values of average inter-string separation (60--160 m) are studied using a Penrose tiling distribution, as illustrated in Fig.~\ref{fig:DetGeoStringSpacing} (a--c). \textcolor{black}{The lower end of this range corresponds to a compact configuration with a total instrumented volume comparable to current-generation detectors such as IceCube or KM3NeT, while the largest spacing results in an instrumented volume roughly twice that envisioned for TRIDENT or IceCube-Gen2. This range thus spans practical detector scales, enabling exploration of the trade-off between effective volume and photon sampling density.}

\paragraph{String height:} The vertical spacing between hDOMs on a single string affects the total height and vertical granularity of the detector. Three vertical spacings (15, 30, 45 m) are considered, shown in Fig.~\ref{fig:DetGeoStringSpacing} (d--f). \textcolor{black}{The maximum tested string height is 900~m, with its top reaching an undersea depth of 2500~m at the TRIDENT site. This was chosen as an upper limit in this study primarily due to engineering constraints during the deployment and stable running of the string array.} 

\paragraph{String clustering:} 

\textcolor{black}{In addition to uniform/non-uniform grids, a cluster layout is also considered by some experiments, such as Baikal-GVD \cite{Baikal-GVD:2025rhg} and P-ONE \cite{P-ONE:2020ljt}. This design is expected to provide dense regions of strings with improved low-energy event reconstruction, combined with sparse inter-cluster arrangements to cover large volumes for high-energy neutrinos. This potentially benefits deep-water instrumentation, allowing for a modular and stepwise construction in challenging deep-water environments. In this work,} the impact of intra-cluster spacing is explored with four values, as illustrated in Fig.~\ref{fig:DetGeoStringSpacing} (g--i), while maintaining the same radius of the full detector volume, \textcolor{black}{which is equal to the largest string spacing tested in the Penrose layout.}

A full list of string and hDOM spacings used in this study are summarised in Table \ref{tab:det_geos}. In all cases, the optical module layout is fixed to the hDOM design proposed in~\cite{Hu:2021jjt}, consisting of 31 3-inch PMTs and 24 SiPM arrays. This work focuses on relative detector layout optimisation; variations in hDOM design have separately been addressed in \cite{hDOM_34inch}.

\begin{table}[!ht]
    \centering
    \caption{Summary of detector geometries considered in this work.}
    \begin{tabular}{|c|c|}
    \hline
       \hline
       \multicolumn{2}{|c|}{String separation ($h_{hDOM} = 30$m)} \\
       \hline
       $<L>$ Average string separation [m] & Detector Volume [km\textsuperscript{3}]\\
       \hline
       60 & 2.2 \\
       80 & 3.9 \\
       100 & 6.1 \\
       120 & 8.8 \\
       140 & 12.0 \\
       160 & 15.6 \\
       \hline\hline
       \multicolumn{2}{|c|}{hDOM vertical separation ($<L> = 100$m) } \\
       \hline
       $h_{hDOM}$ hDOM vertical separation [m] & Detector Volume [km\textsuperscript{3}]\\
       \hline
       15 & 3.1 \\
       30 & 6.1 \\
       45 & 9.2 \\
       \hline\hline
       \multicolumn{2}{|c|}{String clustering (10 strings per cluster, $h_{\text{hDOM}} = 30$m) } \\
       \hline
       $r_{\text{clus}}$ String cluster radius [m] & Detector Volume [km\textsuperscript{3}]\\
       \hline
       80 & \multirow{4}{3em}{15.6} \\
       95 & \\
       115 & \\
       140 & \\
       \hline
    \end{tabular}
    \label{tab:det_geos}
\end{table}

\subsection{Reference TRIDENT Geometry}
\label{sec:reference_geo}

For benchmarking, we define a reference TRIDENT detector configuration with 100 m average inter-string spacing and 30 m vertical hDOM spacing, seen in Fig.~\ref{fig:DetGeoStringSpacing} (b) and (e), following the baseline proposal in~\cite{TRIDENT:2022hql}. This geometry provides a balance between detection volume, photon collection, and practical deployment constraints, and is used as a central comparison point for the optimisation studies that follow.

\section{Reconstruction Performance and Effective Area}

\label{sec:reconstruction}

\begin{figure}[!ht]
    \centering
    \begin{subfigure}{0.45\textwidth}
        \includegraphics[height=5.5cm]{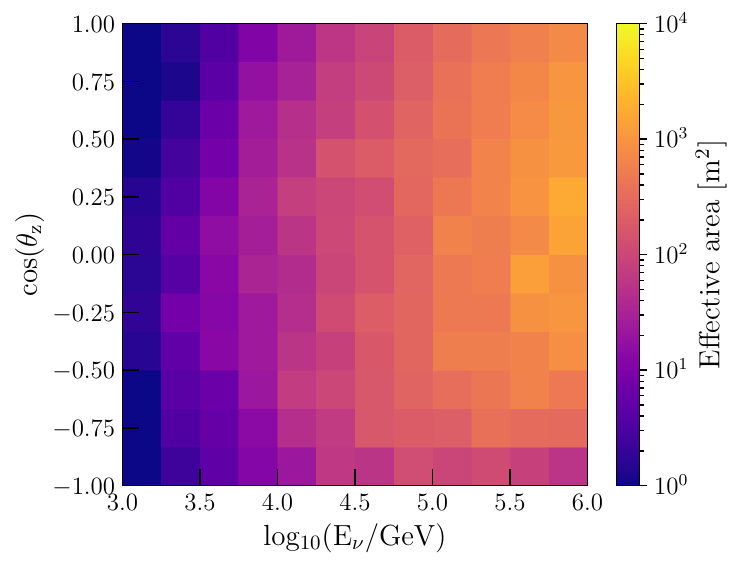}
        \subcaption{}
        \label{fig:track_aeff}
    \end{subfigure}
    \begin{subfigure}{0.45\textwidth}
        \includegraphics[height=5.5cm]{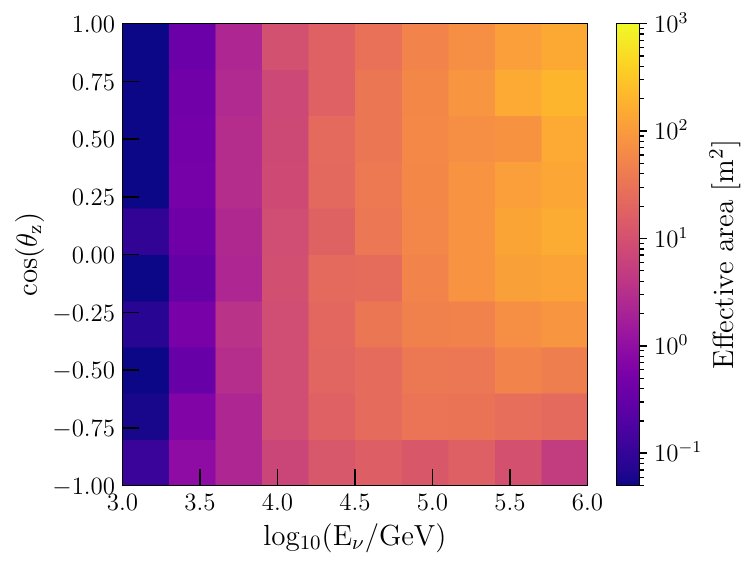}
        \subcaption{}
        \label{fig:cascade_aeff}
    \end{subfigure}
    \vspace{0.5cm}
    \begin{subfigure}{0.45\textwidth}
        \includegraphics[height=5.5cm]{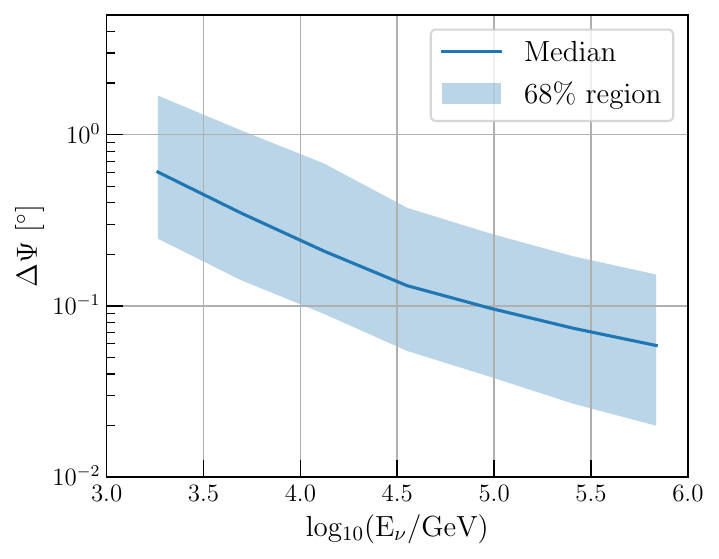}
        \subcaption{}
        \label{fig:track_ar}
    \end{subfigure}
    \begin{subfigure}{0.45\textwidth}
        \includegraphics[height=5.5cm]{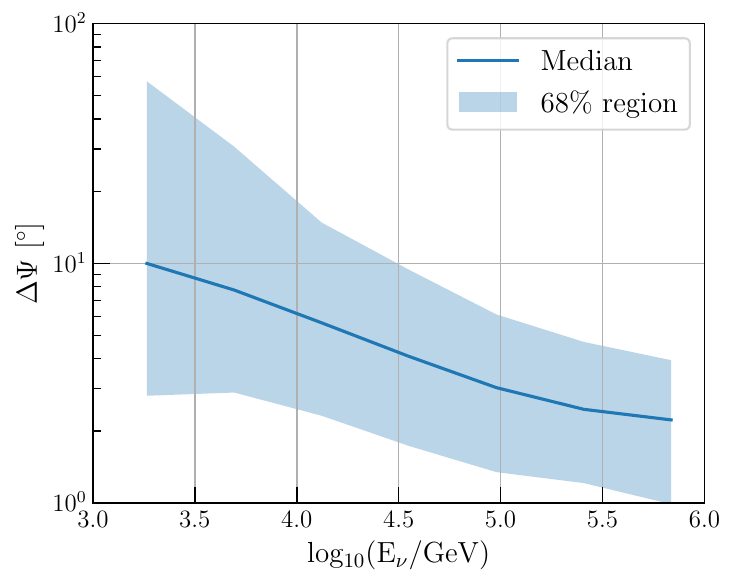}
        \subcaption{}
        \label{fig:cascade_ar}
    \end{subfigure}
    \caption{Effective area (top) and angular resolution (bottom) for $\nu_\mu$ CC track events (left) and $\nu_e$ CC cascade events (right), for the reference TRIDENT geometry, applying the trigger and reconstruction cuts described in section \ref{sec:Aeff}.}
    \label{fig:track_cascade_aeff_ar}
\end{figure}

\textcolor{black}{To assess the performance of each detector configuration, all-flavor neutrino events are simulated and reconstructed. For each layout, the effective area and angular resolution are calculated, providing fundamental measures of the detector’s capabilities. The effective area quantifies the detection efficiency of neutrinos across different flavors and energies, while the angular resolution describes the precision with which the detected neutrino’s incoming direction can be reconstructed. This section describes the calculation of the effective area for tracks and cascades, alongside the angular resolution of detectable events. With these, the point source discovery potential can be calculated and compared for various detector geometries, discussed in detail in Section~\ref{sec:point_source_discovery potential}.}

\subsection{Effective Area for Tracks and Cascades}
\label{sec:Aeff}

The effective area $A_\mathrm{eff}$ converts a given neutrino flux to an observable event rate:
\[ \frac{dN}{dt} = \int d\Omega \int A_\mathrm{eff}(E, \cos \theta) \times \Phi(E) dE \]

Undersea telescopes such as TRIDENT encounter substantial background rates from several sources, including radioactive decays (e.g., $^{40}$K in seawater), radioactivity in hDOM glass, intrinsic PMT dark noise, and atmospheric muons and neutrinos. Dedicated studies combining on-site measurements~\cite{TRIDENT:2022hql,ccsn_icrc2025} and \textsc{Geant4}-based simulations have been used to characterize these contributions, enabling realistic background modeling for triggering and reconstruction algorithms. Effects due to bioluminescence, uncertainties in hDOM positioning, and potential spatial or temporal variations in optical properties are not included in the present study. These are the subject of dedicated calibration studies for TRIDENT. Since the primary aim here is to compare the relative source discovery performance of different detector geometries, the omission of these effects is expected to have only a minor influence on the comparative results. Triggering and reconstruction quality cuts are applied to reduce the influence of background events and improve angular resolution, as described below.

\subsubsection{Event Trigger}
\label{sec:evt_trigger}

A triggering condition of 5 hDOMs is applied~\cite{hDOM_34inch}, where each hDOM is defined as being triggered by having at least 2 PMT hits within 20~ns. This is expected to provide a manageable background rate while maintaining sensitivity to astrophysical neutrinos from $\sim$100 GeV to PeV energies.

\subsubsection{Reconstruction Quality Cuts}
\label{sec:recon_quality}

After the trigger selection, further quality cuts are applied to remove events that would yield poor angular reconstruction, while retaining the majority of well-reconstructed signal events.  
These cuts are applied to all detector geometries considered in this work, where cut  threshold values are individually tuned for each geometry so that they remove the same fraction of events classified as poorly reconstructed, defined further below. 

\paragraph{Edge Cut:} Events with primary energy depositions located too close to the detector's physical boundary are rejected, as they are more likely to suffer from incomplete light collection and biased reconstruction. Edge events are defined with a barycenter position, which corresponds to the hit-weighted average position of all hDOMs detecting photons in the event.  For each geometry, the edge cut corresponds to a cylindrical volume one string spacing inward from the instrumented boundary (or one cluster radius inward for clustered layouts).

\paragraph{Upward Direction Cut (Tracks only):} To suppress the dominant background from down-going atmospheric muons, only reconstructed tracks 10$^{\circ}$ below the detector's horizon are retained.

\paragraph{Track Extension Cut (Tracks only):} Tracks producing light only in a small, localized region of the detector are prone to poor direction reconstruction. To quantify spatial light distribution, we define a hits-weighted standard deviation of hDOM positions: $\sigma_w = \sqrt{\frac{\sum n_i\cdot||<\vec{r}_Q> - \vec{r}_i||^2}{N_{tot}}}$, where $n_i$ is the number of photon hits on the $i^{\mathrm{th}}$ hDOM, $\vec{r}_i$ its position, $N_\text{tot}$ the total number of detected photons, and $\langle \vec{r}_Q \rangle$ the hit barycenter. Events with $\sigma_w$ below a geometry-dependent threshold are removed, as they exhibit insufficient spatial track extension for reliable reconstruction. 
The cut value of $\sigma_w$ applied to each geometry is scaled such that 80\% of events with angular difference between the reconstructed and true neutrino directions exceeding $6^\circ$, are removed. This geometry-dependent cut is used to promote consistent reconstruction quality across configurations.

The resulting effective areas for $\nu_\mu$ CC tracks and $\nu_e$ CC cascades, after application of the trigger and reconstruction quality cuts described above, are shown in Fig.~\ref{fig:track_aeff} and Fig.~\ref{fig:cascade_aeff}, respectively.

\subsection{Direction Reconstruction}

\subsubsection{Direction Reconstruction for Tracks}

For events passing the trigger and quality cuts described in Section~\ref{sec:Aeff}, a likelihood-based direction reconstruction algorithm is applied. The method utilises a multi-dimensional probability distribution function (PDF) produced from a large number of simulated track events within the TRIDENT detector. Two of the three PDF dimensions consist of a characteristic distance-dependent time residual, reflecting the detected photon arrival time minus the time of flight from a muon track (emitted at the Cherenkov angle in water), and the muon production time. This time residual distribution is quantified as a function of the distance of a track from a given hDOM, assuming again that light is emitted from the track at the characteristic Cherenkov angle.
The third and final dimension is the angle of a given PMT orientation on an hDOM to the assumed photon production position, allowing for positive weighting of PMTs facing the track event. Fig.~\ref{fig:track_ar} shows the angular resolution in the reference TRIDENT geometry for track events, following the application of the cuts discussed above.

\subsubsection{Direction Reconstruction for Cascades}

The vertex and direction reconstruction method for cascade-like events also uses a maximum likelihood approach. An initial vertex position is reconstructed using a distance-independent time residual PDF, treating the cascade as a point-like isotropic light source. Then, both direction and vertex are optimised using a Poisson-based likelihood:
\[ \log \mathcal{L} = \sum (k_i \log \mu_i - \mu_i) \]
where $k_i$ and $\mu_i$ are the observed and expected photon hits on individual PMTs, parameterized again in three dimensions, namely: the distance between the cascade central vertex and the hDOM, the angle between the cascade direction and the hDOM, and finally the angle between a hit PMT position and the assumed cascade vertex. Fig.~\ref{fig:cascade_ar} shows the angular resolution for cascade events due to $\nu_e$ CC interactions in the reference TRIDENT geometry, after triggering and quality cuts.

\begin{figure}[!ht]
    \centering
    \begin{subfigure}{0.4\textwidth}
        \centering
        \includegraphics[height=5cm]{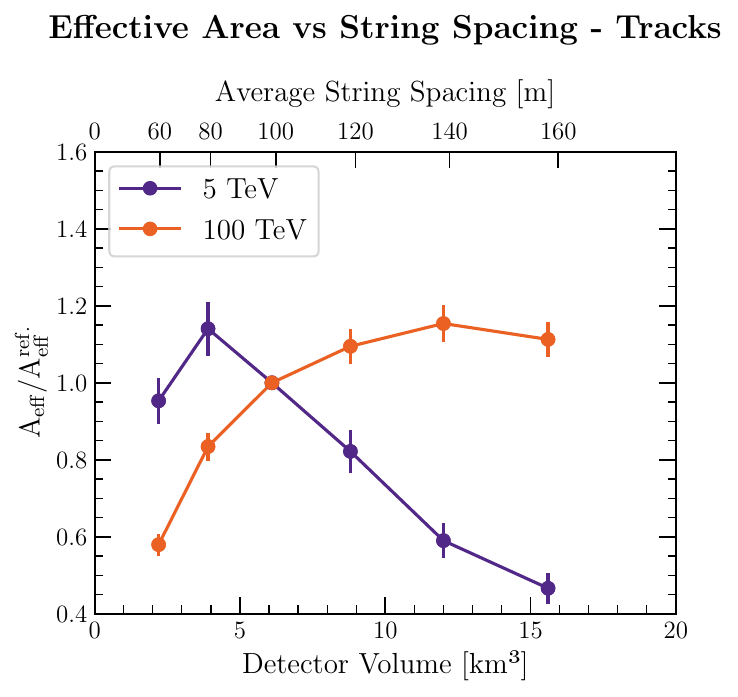}
    \end{subfigure}
    \begin{subfigure}{0.4\textwidth}
        \centering
        \includegraphics[height=5cm]{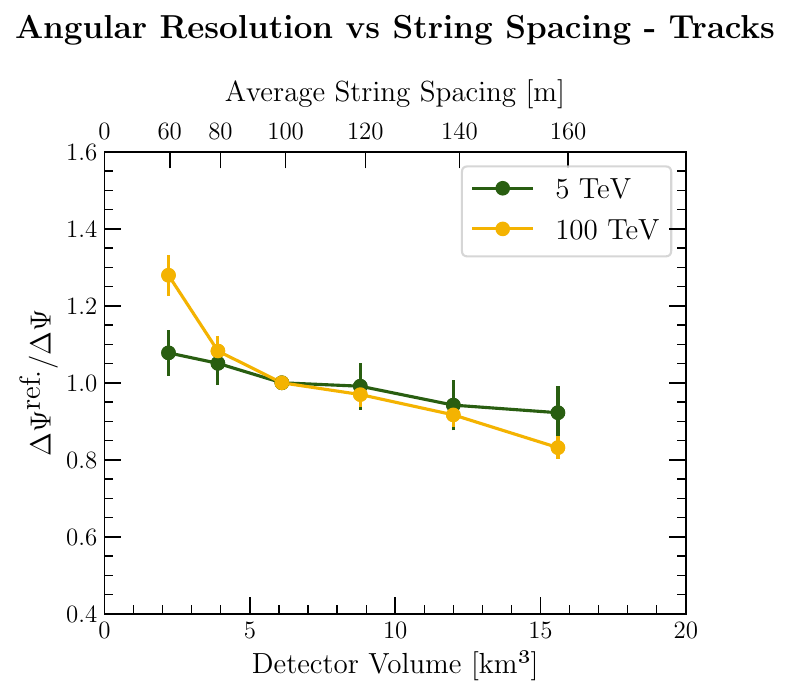}
    \end{subfigure}

    \vspace{0.5cm} 

    \begin{subfigure}{0.4\textwidth}
        \centering
        \includegraphics[height=5cm]{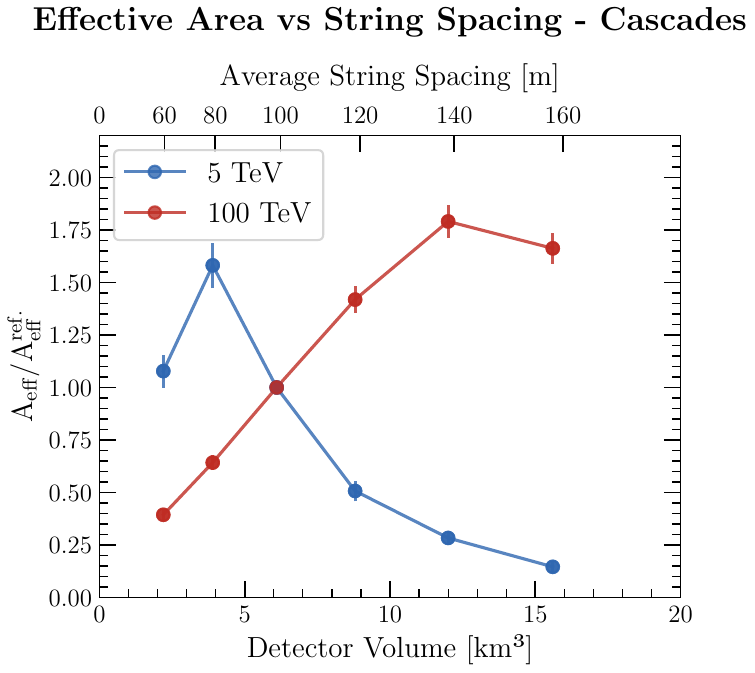}
    \end{subfigure}
    \begin{subfigure}{0.4\textwidth}
        \centering
        \includegraphics[height=5cm]{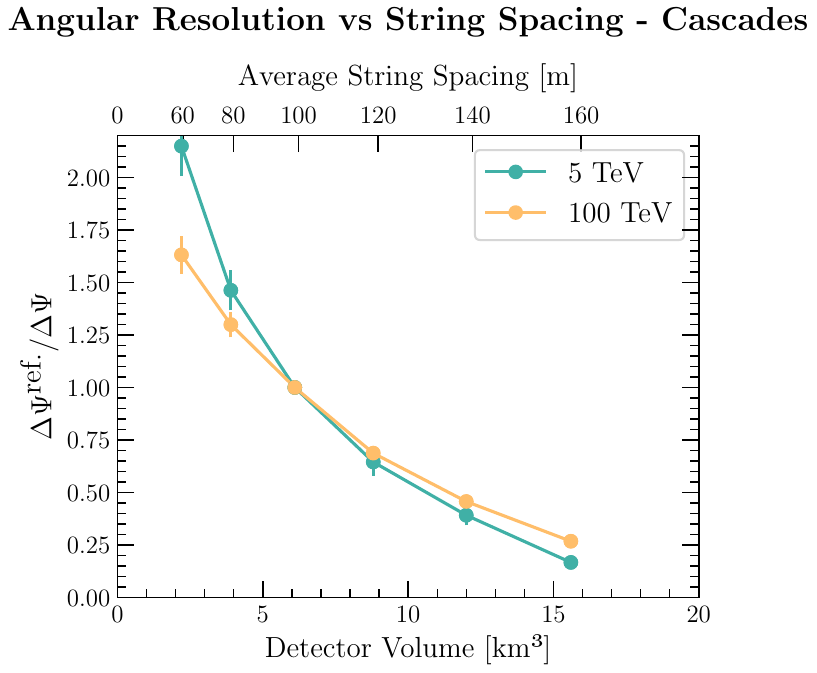}
    \end{subfigure}
    \caption{Ratios of the effective area for $\nu_{\mu}$-CC track events and $\nu_e$-CC cascade events in (left top and left bottom, respectively) with respect to the reference TRIDENT geometry, shown as a function of string spacing and consequently detector volume. Angular resolution ratios are similarly shown (right). Uncertainty bars reflect the simulated event statistics at a given energy. Values larger than unity in all plots indicate improved performance for the tested layout compared to the reference geometry.}
    \label{fig:string_spacing_track_cascade}
\end{figure}

\begin{figure}[!ht]
    \centering
    \begin{subfigure}{0.4\textwidth}
        \centering
        \includegraphics[height=5cm]{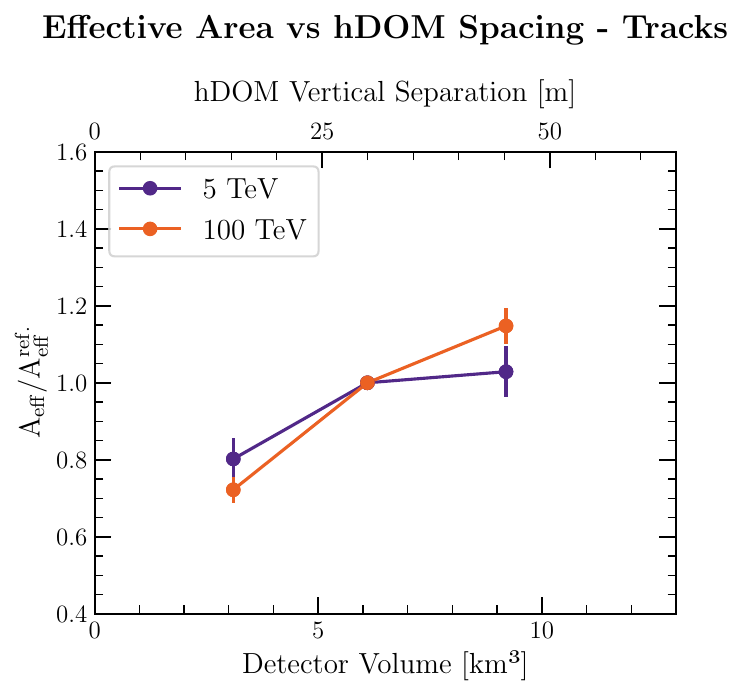}
    \end{subfigure}
    \begin{subfigure}{0.4\textwidth}
        \centering
        \includegraphics[height=5cm]{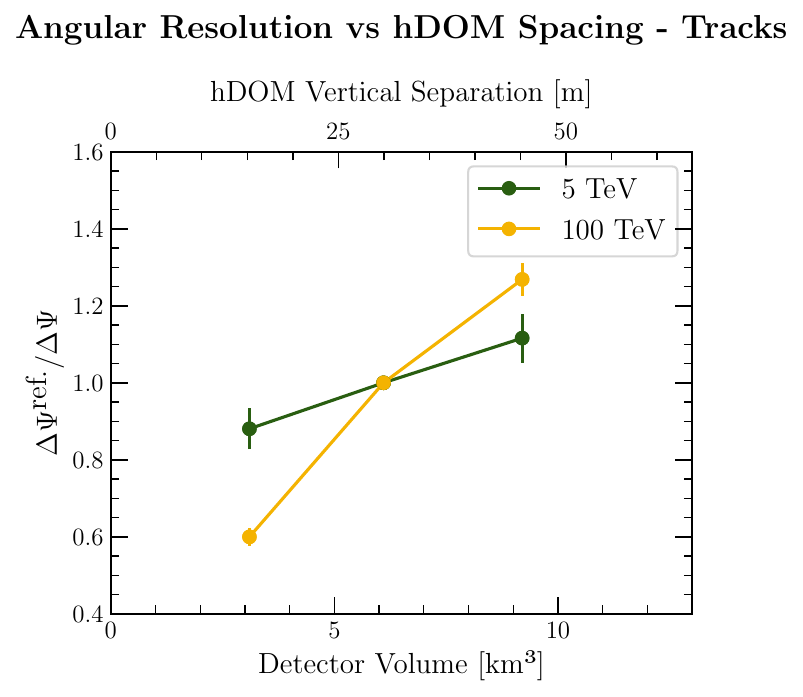}
    \end{subfigure}
    \label{fig:vertical _tracks}

    \vspace{0.5cm} 
    
    \begin{subfigure}{0.4\textwidth}
        \centering
        \includegraphics[height=5cm]{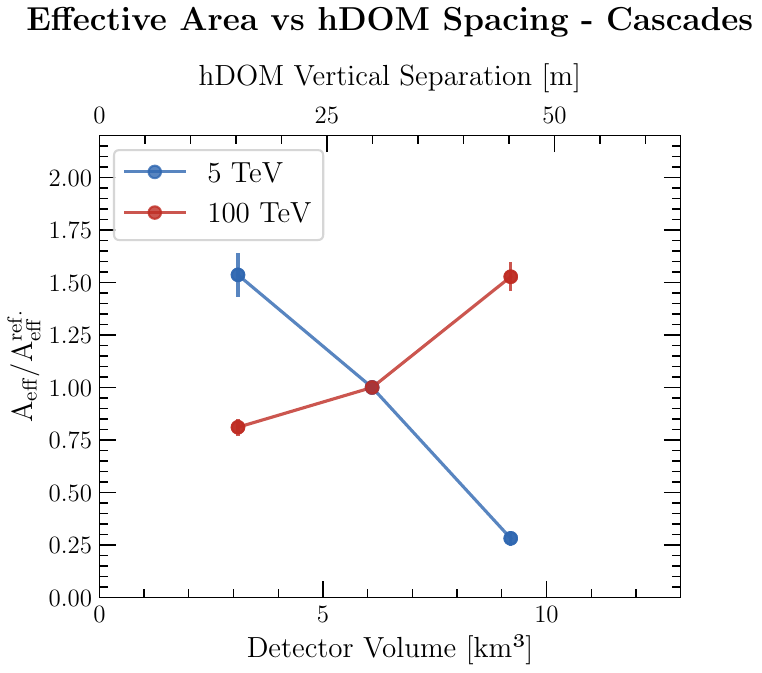}
    \end{subfigure}
    \begin{subfigure}{0.4\textwidth}
        \centering
        \includegraphics[height=5cm]{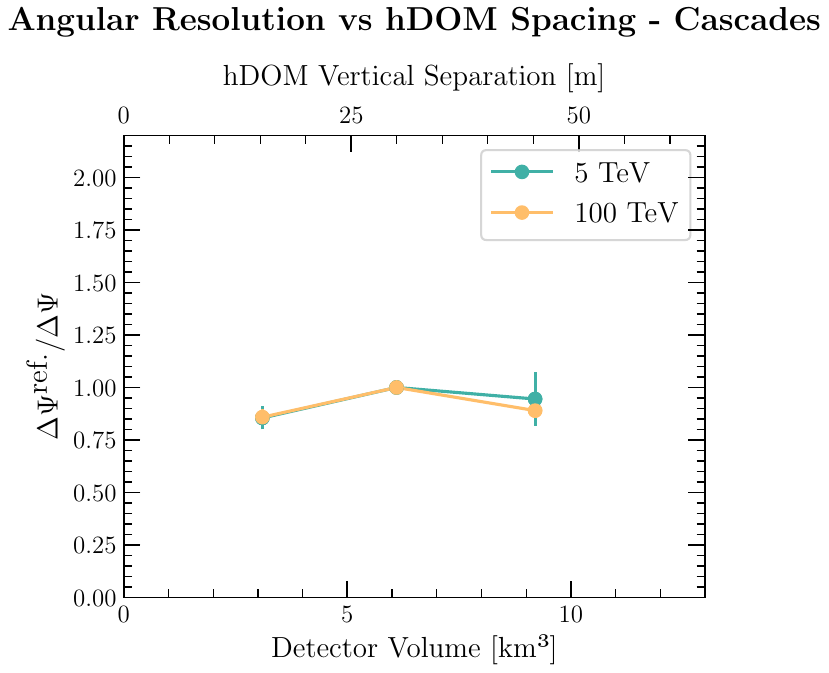}
    \end{subfigure}
    \caption{Ratios of the effective area for $\nu_{\mu}$-CC track events and $\nu_e$-CC cascade events in (left) with respect to the reference TRIDENT geometry, shown as a function of detector volume, varying the string height. Angular resolution ratios are similarly shown (right). Values larger than unity in all plots indicate improved performance for the tested layout compared to the reference geometry.}
    \label{fig:vertical_tracks_cascades}
\end{figure}

\begin{figure}[!ht]
    \centering
    \begin{subfigure}{0.4\textwidth}
    \includegraphics[height=5cm]{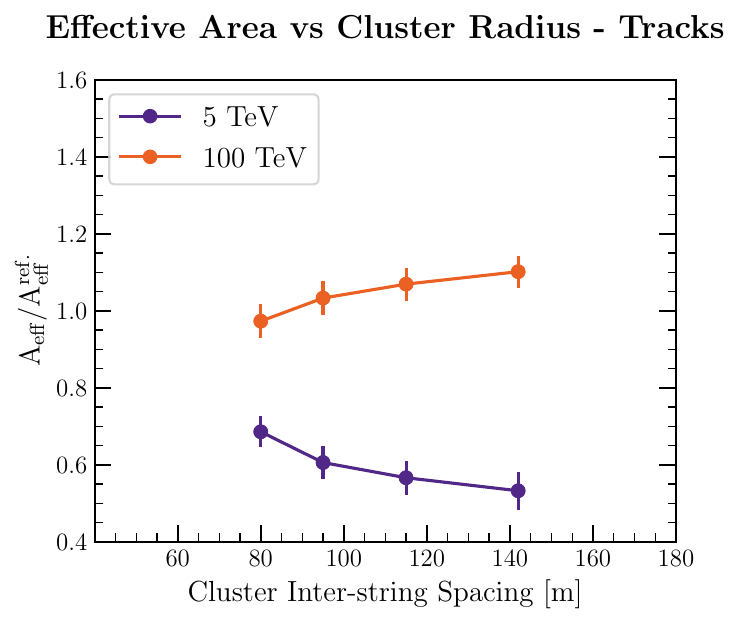}
    \end{subfigure}
    \begin{subfigure}{0.4\textwidth}
    \includegraphics[height=5cm]{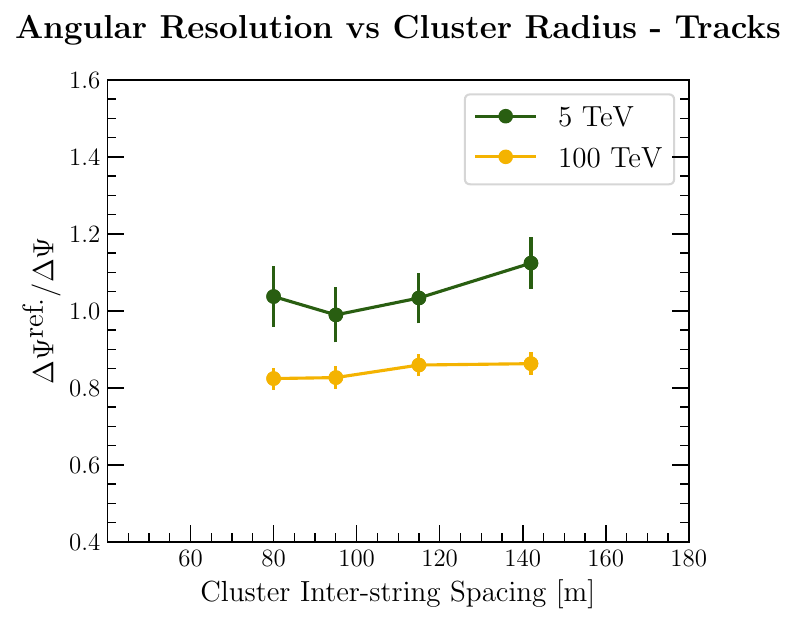}
    \end{subfigure}

    \vspace{0.5cm} 
    
    \begin{subfigure}{0.4\textwidth}
    \includegraphics[height=5cm]{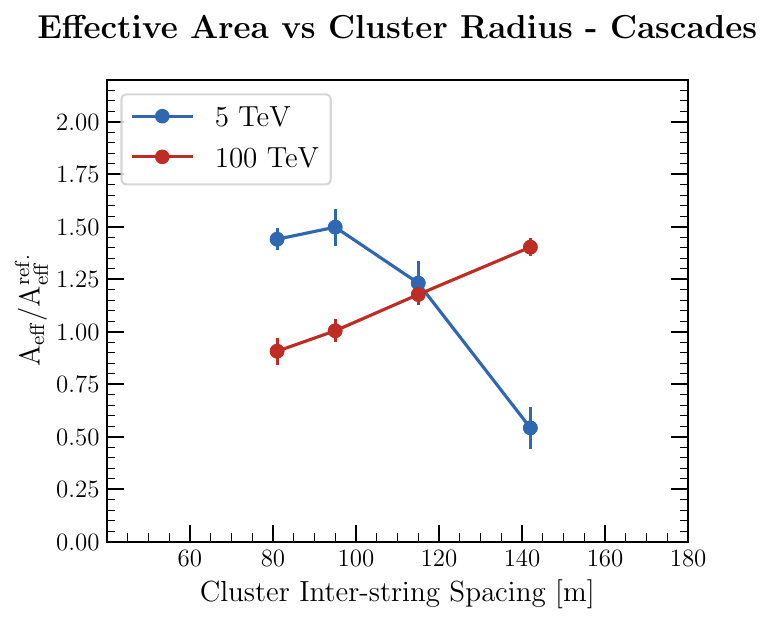}
    \end{subfigure}
    \begin{subfigure}{0.4\textwidth}
    \includegraphics[height=5cm]{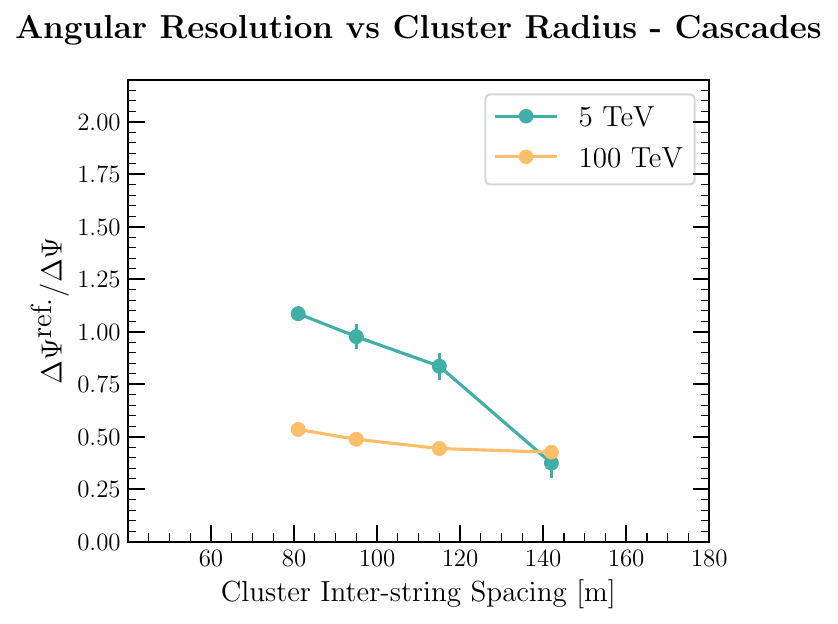}
    \end{subfigure}
        \caption{Ratios of the effective area for $\nu_{\mu}$-CC track and $\nu_e$-CC cascade events in (left) with respect to the reference TRIDENT geometry, shown as a function of the radius of each string cluster, assuming a fixed total detector instrumented volume of 15.6 km\textsuperscript{3}, summarized in section \ref{sec:reconstruction}. Angular resolution ratios are similarly shown (right). Values larger than unity in all plots indicate improved performance for the tested layout compared to the reference geometry.}
    \label{fig:clustering_tracks_cascades}
\end{figure}

\subsection{Performance Comparisons for Various Detector Geometries}

\label{sec:geom_dependence}

Figures~\ref{fig:string_spacing_track_cascade}--\ref{fig:clustering_tracks_cascades} show the ratios of effective area and angular resolution for $\nu_\mu$-CC tracks and $\nu_e$-CC cascades, relative to the reference TRIDENT geometry, for 5~TeV and 100~TeV neutrinos. These aim to represent low- and high-energy regimes, highlighting differences in detection efficiency and reconstruction performance.

\paragraph{String spacing:}
Fig. \ref{fig:string_spacing_track_cascade} shows that increasing string spacing (and thus detector volume) boosts high-energy track effective area, but grows sub-linearly, as km-scale muons extend beyond the instrumented region. At low energy, sparse layouts lose efficiency as many tracks fail to trigger or reconstruct well. For cascades, high-energy effective area scales with volume until the widest spacings reduce photon collection; at low energy, efficiency peaks for slightly smaller volumes than the reference. Angular resolution worsens with sparser layouts for both event types due to reduced photon statistics. However, the degradation is less steep for tracks, since extended muon paths can still produce useful distant photon detections along their trajectory, providing additional geometric constraints on direction. Cascades, being more point-like, lack this benefit and are therefore more sensitive to reduced module density.

\paragraph{String height:}
Fig. \ref{fig:vertical_tracks_cascades} demonstrates that larger vertical hDOM separations (greater total string height) increase effective area for tracks at all energies, with minimal photon loss. High-energy cascades also benefit, but low-energy cascades see reduced efficiency. Track angular resolution improves modestly with height, thanks to longer vertical lever arms, while cascades are largely unaffected.

\paragraph{String clustering:}
As seen in Fig. \ref{fig:clustering_tracks_cascades}, reducing clustering (more uniform layouts) improves high-energy track and cascade effective area by increasing coverage. Low-energy tracks are strongly penalized by clustering, as many pass through gaps without triggering. Low-energy cascades gain efficiency from tighter clustering, but high-energy cascades lose efficiency. For angular resolution, low-energy events can benefit slightly from clustering, while high-energy events, particularly cascades, tend to worsen with greater cluster separation.

\vspace{0.5cm}

Overall, these trends reflect a balance between instrumented volume and photon detection efficiency: larger, sparser detectors capture more high-energy events, while dense configurations are essential for reconstruction performance across a wide energy range. In the next section, detection rates and reconstruction quality are quantitatively combined to determine the point source discovery potential and guide the final geometry optimisation.

\section{Point Source Discovery Potential}
\label{sec:point_source_discovery potential}

The combination of a detector’s effective area and angular resolution determines both the expected astrophysical signal rate and the precision with which its origin in the sky can be reconstructed. To compare TRIDENT design options, we define the point source discovery potential as a figure of merit, representing the flux needed for a $5\sigma$ discovery of a steady point source within a given observation time. This metric naturally balances the competing effects of event statistics and directional precision, and is evaluated using neutrinos of all flavours, a primary physics goal of TRIDENT.

\subsection{Discovery Potential}
The analysis assumes astrophysical neutrino fluxes following a power-law spectrum:
\[
\frac{d\Phi_{\nu}}{dE} = \phi_0 \left(\frac{E}{1~\mathrm{TeV}}\right)^{-\gamma},
\]
with $\gamma=2$ (hard spectrum) and $\gamma=3$ (soft spectrum) as representative cases. Simulated samples include $\nu_{e,\mu,\tau}$ and corresponding antineutrinos, with CC and NC interactions accounting for their relative cross-sections. Event rates for each flavour and interaction types are derived from the high-statistics $\nu_e$-CC and $\nu_\mu$-CC simulations via energy-dependent scaling built using separate full-flavour simulation samples of CC and NC interactions.

\vspace{0.5cm}

Two primary background components are included in discovery potential calculations:
\begin{itemize}
    \item \textbf{Atmospheric neutrinos}: the flux of which is taken from the \texttt{Daemonflux} package~\cite{Yanez:2023lsy}, including conventional and prompt components tuned with high-energy muon data.
    \item \textbf{Diffuse astrophysical neutrinos}: isotropic, equal-flavour flux at Earth with $\gamma=2.37$ and $\phi_0=1.44$ as measured by IceCube~\cite{ICDiffuseAstro}.
\end{itemize}

Angular spread PDFs are generated for signal and background and combined into Asimov datasets. An extended binned likelihood fit is performed, where the fitted source normalisation $\phi_0$ corresponds to the $5\sigma$ discovery threshold. The figure of merit for geometry comparison is the time required to reach the $5\sigma$ level for a fixed reference source flux.

\subsection{Comparison of Point Source Discovery Potential}

\begin{figure}[!ht]
    \centering
    \begin{subfigure}{0.45\textwidth}
        \centering
        \includegraphics[height=5cm]{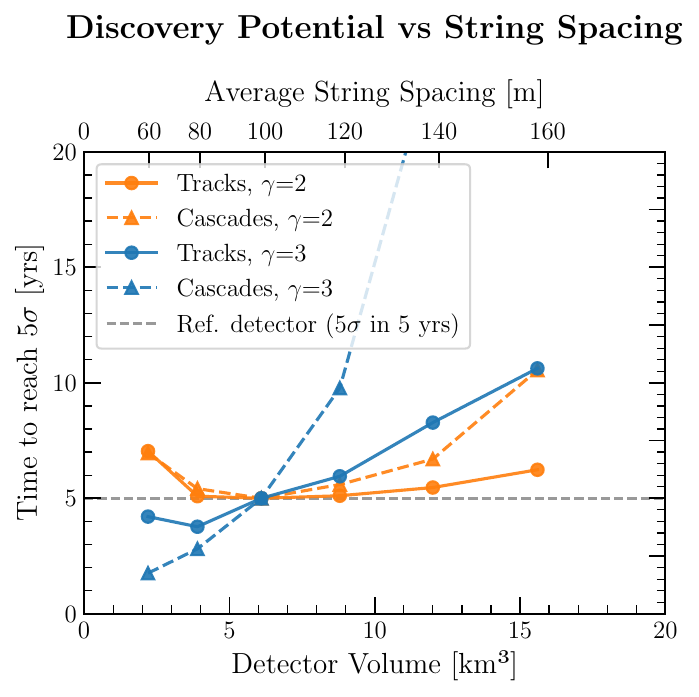}
        \subcaption{}
    \end{subfigure}
    \begin{subfigure}{0.45\textwidth}
        \centering
        \includegraphics[height=5cm]{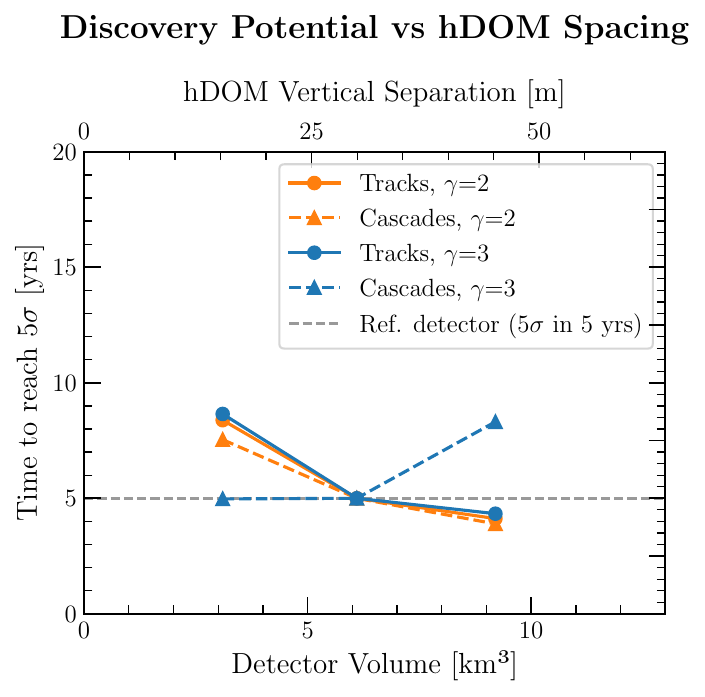}
        \subcaption{}
    \end{subfigure}    
    \begin{subfigure}{0.45\textwidth}
        \centering
        \includegraphics[height=5cm]{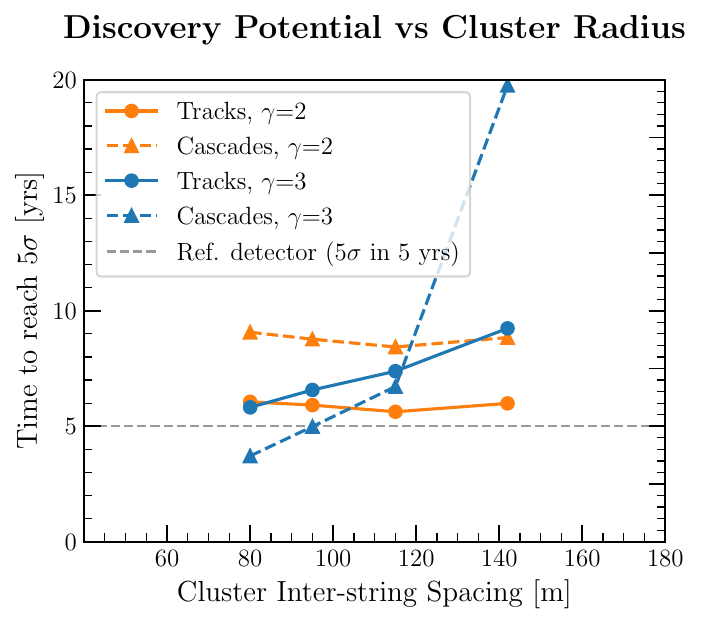}
        \subcaption{}
    \end{subfigure}
    \begin{subfigure}{0.45\textwidth}
        \centering
        \includegraphics[height=5cm]{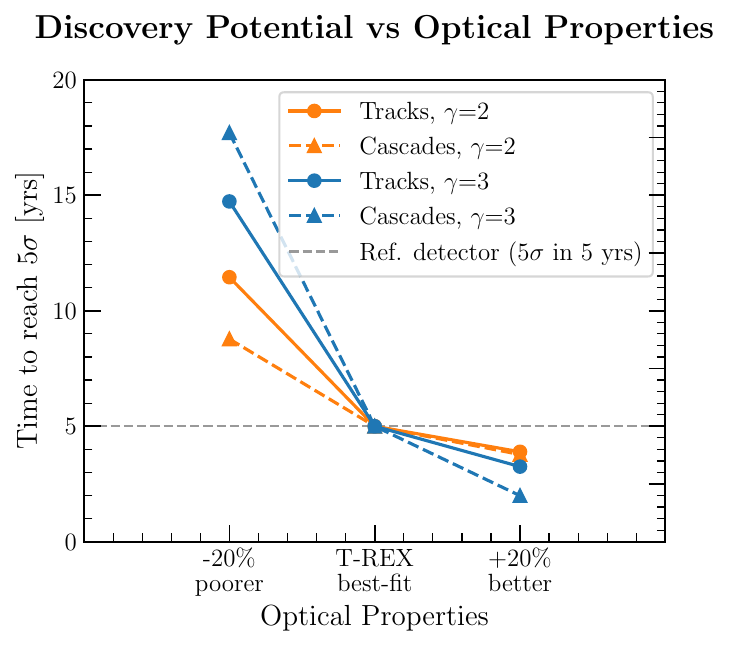}
        \subcaption{}
    \end{subfigure}
    \caption{Relative time to reach the same $5\sigma$ discovery potential as the reference TRIDENT geometry for (a) string spacing, (b) string height, (c) string clustering, and (d) $\pm 20\%$ changes in seawater attenuation length. Lower values indicate better performance.}
    \label{fig:DP_Time_vs_Geo}
\end{figure}

Fig.~\ref{fig:DP_Time_vs_Geo} shows the relative time required for each detector geometry to reach the same $5\sigma$ discovery flux as the reference TRIDENT layout, for both track and cascade events and for hard ($\gamma=2$) and soft ($\gamma=3$) source spectra. The reference geometry is normalised to a 5-year discovery time; highlighting the shorter/longer time needed for a given detector geometry to reach the same level of significance for the same source. Here, the discovery potential was evaluated assuming a neutrino source located at declination $\sin(\delta)=0$, for all detector geometries. While discovery potentials were calculated for all source declinations, the choice of source location was not found to significantly impact the relative discovery time of the geometries considered.

\paragraph{String Spacing:}  
In Fig.~\ref{fig:DP_Time_vs_Geo}(a), reducing the string spacing below the reference improves (reduces) discovery time for soft-spectrum cascades (blue-dashed) by up to a factor of 2, and approximately $\sim20\%$ for tracks (blue-solid) reflecting their reliance on high photon statistics for precise reconstruction. However, this same denser layout slightly worsens performance for hard-spectrum tracks and cascades (black-solid and dashed, respectively), where the smaller instrumented volume reduces the number of high-energy events despite better photon statistics. Increasing the string spacing above the reference is seen to worsen discovery times for both tracks and cascades, assuming either hard or soft energy spectra. Despite increases in acceptance for high energy neutrinos, larger spacings cause loss of angular resolution (especially for cascades) and losses in low energy neutrino sensitivity, causing discovery times to worsen rapidly. Tracks are seen to degrade more gradually than cascades with increased string spacings, since extended muon paths still provide useful directional constraints from distant photon detections.

\paragraph{String Height:}  
Fig.~\ref{fig:DP_Time_vs_Geo}(b) shows that modest increases in string height generally improve sensitivity for all cases (excluding soft spectra cascades), shortening discovery times by $\sim$10--20\%. Reduced vertical photon coverage shows degradation of both effective area and reconstruction, leading to extended discovery times, highlighting the benefit of taller strings. Engineering constraints such as susceptibility to sea currents may ultimately set practical limits on the string height optimisation.

\paragraph{String Clustering:}  
As seen in Fig.~\ref{fig:DP_Time_vs_Geo}(c), clustering strings together worsens sensitivity for most channels, especially for hard-spectrum sources. For the most clustered layouts, discovery times worsen for tracks and cascades by up to $\sim$20\% and $\sim$80\%, respectively. An exception occurs for soft-spectrum cascades, which show steady improvement with increased clustering; here, the locally dense hDOM coverage boosts photon statistics enough to outweigh the loss in event rate acceptance.

\paragraph{Optical Properties:}  
While the results above compare purely geometrical configurations, the performance of any deep-sea neutrino telescope is  dependent on the optical properties of the surrounding water — a factor not varied in the previous sections. The attenuation length, which quantifies the combined effects of photon absorption and scattering, directly controls the photon yield reaching the hDOMs for both tracks and cascades.

To assess its influence, the reference TRIDENT geometry was simulated with $\pm20\%$ variations in attenuation length (with reconstruction pdfs tuned in each case), with discovery times shown in Fig.~\ref{fig:DP_Time_vs_Geo}(d). The resulting changes in discovery potential are striking: a 20\% degradation in attenuation length can lengthen the time to discovery by factors of $\sim$2--4 times, penalties equal or larger than the previous geometrical design changes considered. Conversely, modest improvements in optical properties yield benefits across all source types.

This has two important consequences. First, the ranking of detector geometries can shift significantly if optical properties deviate from nominal expectations, implying that design decisions made under optimistic assumptions could be suboptimal in real conditions. This highlights the importance of initial string deployment in TRIDENT Phase-I, to carefully asses the impact of the water optical properties, and use the results to \textit{lead} the design of the full TRIDENT layout. Second, it emphasises the need for continuous monitoring of water clarity \cite{T-REX:2024qfj} and, where possible, mitigation strategies (e.g. dense and sparse hDOM and string spacings, and adaptive trigger settings) to preserve sensitivity over the detector’s lifetime. In short, the optical environment is not merely a boundary condition, but a design driver that must be considered alongside the geometry when planning TRIDENT’s final layout.

\section{Conclusion}
\label{sec:conclusion}


\textcolor{black}{We have presented a performance study of the design of next-generation deep-sea neutrino telescopes, using TRIDENT as a case study. We quantified how detector geometry and the optical properties of the seawater influence effective area, angular resolution, considering both track and cascade-like events. Building upon these results, we evaluated the discovery potential for steady astrophysical point sources. Full-chain simulations with site-specific optical parameters and realistic atmospheric and diffuse backgrounds were used to evaluate the $5\sigma$ discovery time for different detector configurations.}


\textcolor{black}{Varying detector geometry has a clear influence on discovery performance across different energy ranges and event types. Wider string spacings enhance sensitivity to the highest energy neutrinos by increasing the instrumented volume, but this benefit quickly saturates and comes at the cost of reduced photon statistics and degraded angular resolution. Conversely, denser layouts improve low-energy performance but limit acceptance at the highest energies. Increasing the vertical extent of the strings offers moderate performance gains, though it must be balanced against engineering and stability constraints in deep-sea conditions. While clustering strings can simplify deployment and maintenance, the empty volumes between string clusters are found to diminish point-source sensitivity for almost all channels, promoting a more uniform string layout. Taken together, these results show that the reference TRIDENT geometry represents a balanced compromise, providing robust sensitivity to both hard and soft source spectra across a wide range of energies.}

Beyond geometry choices, we find that the optical properties of the seawater can have an impact on discovery potential equal to, or greater than, the geometrical changes studied. A 20\% reduction in attenuation length can increase discovery times by factors of two to four, potentially overturning the relative ranking of geometry options. This underscores that environmental conditions are not just a boundary constraint but an active design driver. Early in-situ measurements with Phase-I deployments will therefore be essential to anchor the final TRIDENT layout to realistic optical conditions, with continuous monitoring throughout operations.


\textcolor{black}{In summary, detector design should be primarily guided by the physics goals of the experiment. This study demonstrates that, for the purpose of efficient source discovery with all neutrino flavors, larger is not inherently better for deep-sea neutrino telescopes, particularly when photon propagation distances are limited. Based on the ambient optical properties measured at the TRIDENT site, the reference TRIDENT geometry is found to provide balanced performance for both track and cascade performance across a wide energy range, where a larger, sparser detector is found to perform more poorly in all categories. The optical properties of the detector medium also strongly influence detector performance and therefore its optimisation. If the optical properties measured in the Mediterranean Sea \cite{Yepes-Ramirez:2013jna} are assumed, an optimal design for the TRIDENT geometry are expected to shift toward a larger, sparser detector configuration. While constructing separated clusters of strings can simplify deployment and maintenance in the deep ocean, and can cover a large total instrumented volume, the performance does not scale directly with this volume. Empty spaces between clusters reduce photon coverage, while the larger fraction of edge events degrades both reconstruction quality and event classification. Future optimisation of the TRIDENT detector will integrate additional physics drivers in parallel, including sensitivity to the neutrino flavor ratio, building on TRIDENT’s existing work on tau neutrino identification \cite{wei_thesis}. Careful inclusion of such multi-channel performance metrics into the design process will ensure that the planned next-generation detector delivers maximum scientific return in both source discovery and flavor-sensitive astrophysics.}

\section*{Acknowledgements}

This work has been supported by the Ministry of Science and Technology of China under Grant No.~2022YFA1605500 and the Office of Science and Technology of the Shanghai Municipal Government under Grant No. 22JC1410100. I. Morton-Blake also acknowledges the National Natural Science Foundation of China Grant No. 12350410499 and the Kuan-Cheng Wang Education Foundation.

\bibliographystyle{unsrturl}
\bibliography{ref}

\end{document}